\documentclass[10pt,journal,twoside,web]{ieeecolor}
\usepackage{tmi2}
\usepackage{cite}
\usepackage{framed,multirow}
\usepackage{comment}
\usepackage{latexsym}
\usepackage{amsmath,amssymb,amsfonts}
\usepackage{algorithmic}
\usepackage{graphicx}
\usepackage{textcomp}
\usepackage{stackengine}

\usepackage{url}
\usepackage{xcolor}
\usepackage{subfig}
\usepackage{tikz}
\usetikzlibrary{matrix,calc}
\usepackage[colorlinks]{hyperref}
\usepackage[linesnumbered,ruled,vlined]{algorithm2e}
\usepackage{amsmath}
\usepackage{framed,multirow}
\usepackage{amssymb}

\usepackage{latexsym}
\usepackage{lipsum}
\hypersetup{
  colorlinks,
  citecolor=teal,
  linkcolor=blue,
  urlcolor=blue}
\definecolor{newcolor}{rgb}{.8,.349,.1}
\definecolor{brightcerulean}{rgb}{0.11, 0.62, 0.74}

\renewcommand{\arraystretch}{1.4}
\usepackage[font=footnotesize,justification=justified,belowskip=2pt,aboveskip=2pt]{caption}
\SetKwInput{KwInput}{Input}                
\SetKwInput{KwOutput}{Output}              

\makeatletter
\IEEEtriggercmd{\reset@font\normalfont\fontsize{7.9pt}{8.40pt}\selectfont}
\makeatother
\IEEEtriggeratref{1}

\begin{document}
\title{DenoMamba: A fused state-space model \\for low-dose CT denoising}
\author{\c{S}aban \"{O}zt\"{u}rk$^*$, O\u{g}uz Can Duran, and Tolga \c{C}ukur, \IEEEmembership{Senior Member} \vspace{-1.3cm}
\\
\thanks{This study was supported by TUBA GEBIP 2015 and BAGEP 2017 fellowships awarded to T. \c{C}ukur (Corresponding author: \c{S}aban \"{O}zt\"{u}rk, saban.ozturk@hbv.edu.tr).}
\thanks{Authors are with the Dept. of Electrical-Electronics Engineering and National Magnetic Resonance Research Center (UMRAM), Bilkent University, Ankara, Turkey, 06800. \c{S}. \"{O}zt\"{u}rk is also with the Ankara Haci Bayram Veli University, Ankara, Turkey. T. \c{C}ukur is also with the Neuroscience Graduate Program Bilkent University, Ankara, Turkey, 06800. }
}

\maketitle
\begin{abstract}
Low-dose computed tomography (LDCT) lowers potential risks linked to radiation exposure while relying on advanced denoising algorithms to maintain diagnostic image quality. The reigning paradigm in LDCT denoising is based on neural network models that learn image priors to separate noise patterns evoked by dose reduction from underlying tissue signals. Naturally, the fidelity of these priors depend on the underlying model's ability to capture the broad range of contextual features present in CT images. Earlier convolutional models are adept at capturing short-range spatial context, but their limited receptive fields reduce sensitivity to interactions over longer distances. Although transformers help improve sensitivity to long-range context, the native complexity of self-attention operators can elicit a compromise in local precision. To mitigate these limitations, here we introduce a novel denoising method based on state-space modeling, DenoMamba, that effectively captures both short- and long-range context in medical images. Following an hourglass architecture with encoder-decoder stages, DenoMamba employs a spatial state-space modeling (SSM) module to encode spatial context and a novel channel SSM module equipped with a secondary gated convolution network to encode latent features of channel context at each stage. Feature maps from the two modules are then consolidated with low-level input features via a convolution fusion module (CFM). Comprehensive experiments on LDCT datasets with 25\% and 10\% dose reduction demonstrate that DenoMamba outperforms state-of-the-art denoisers based on convolutional, transformer and SSM backbones with average improvements of 1.6dB PSNR, 1.7\% SSIM, and 2.6\% RMSE in image quality.
\end{abstract}


\vspace{-0.3cm}
\bstctlcite{IEEEexample:BSTcontrol}

\section{Introduction}
A cornerstone in modern medical imaging, CT irradiates the body with a beam of X-rays to furnish detailed cross-sectional views of anatomy \cite{10385173}. Unlike conventional radiography, CT relies on acquisition of multiple snapshots as the X-ray beam is rotated around the body, causing substantially elevated exposure to ionizing radiation with potential risks including cancer \cite{10543110}. A mainstream approach to alleviate these health risks involves CT protocols that cap the tube current or exposure time to lower the number of incident photons and thereby the radiation dose \cite{LEI2020101628}. However, as the signal-to-noise ratio (SNR) scales with the number of incident photons, dose reduction inevitably increases the noise component in CT images, significantly degrading image quality and potentially obscuring diagnostic features. Consequently, development of effective denoising methods is imperative to maintaining the diagnostic utility of LDCT images acquired under high levels of dose reduction \cite{9963593}. 

In recent years, deep learning models have superseded traditional approaches in LCDT denoising \cite{89c2e116d9a54a47a4190c1c9ce193b8, 6909762}, given their improved adaptation to the distribution of imaging data \cite{10535288, 10480580}. These models hierarchically process input images across many network stages, wherein multiple sets of latent feature maps are extracted at each stage that encapsulate different image attributes (e.g., edges, textures) in separate feature channels. As tissues can be distributed across broad spatial clusters in anatomical cross-sections \cite{adam2014grainger} and measurement noise variance scales with the intensity of tissue signals \cite{10472025}, latent feature maps of CT images exhibit significant spatial dependencies over short- to long-range distances \cite{ellison2012neuropathology,Wang_2023}. Furthermore, as the network depth increases, higher-levels of latent features are extracted that also manifest strong dependencies across the channel dimension due to overlapping or complementary information. In turn, the success of a denoising model in separating noise from tissue signals depends on its ability to discern idiosyncratic patterns of spatial and channel context in latent feature maps of CT images \cite{TransCT}. 

Earlier studies in learning-based LDCT denoising have predominantly employed convolutional neural network (CNN) models to process LDCT images \cite{https://doi.org/10.1002/mp.12344,8946589,2017arXiv170200288C,9320928,9669891,li2023multi}. CNN models employ compact convolution operators for image processing, and hence perform local filtering driven by spatial distance between image pixels. This locality bias yields linear model complexity with respect to image dimensions, and offers high expressiveness for local contextual features that are critical in delineating detailed tissue structure \cite{8718010,LI2024106329}. However, it inevitably restricts sensitivity to long-range contextual features in CT images, whether instigated across the spatial or channel dimensions \cite{resvit}. Therefore, CNNs can suffer from poor denoising performance especially near regions of heterogeneous tissue composition, where understanding spatial and channel dependencies of latent feature maps can be crucial for distinguishing signal from noise.

\begin{figure*}[!t]
\centerline{\includegraphics[width=0.9\textwidth]{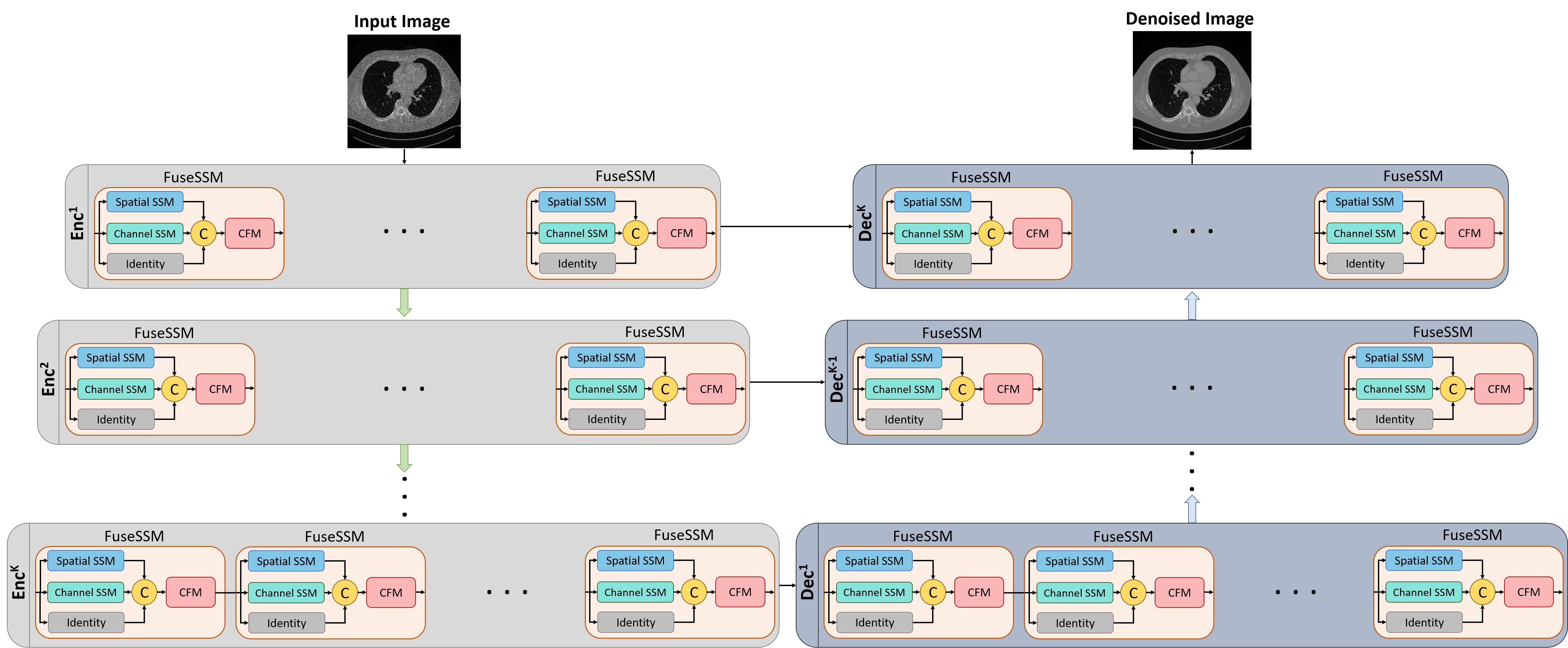}}
\caption{Overall architecture of DenoMamba. The proposed model comprises encoder-decoder stages that are residually connected with long skip connections. In the encoder stages, input feature maps are projected through cascaded FuseSSM blocks, and spatially downsampled while the channel dimensionality is increased. In the decoder stages, input feature maps are back-projected through cascaded FuseSSM blocks, and spatially upsampled while the channel dimensionality is reduced. The proposed FuseSSM blocks use a spatial SSM module to extract spatial context, a novel channel SSM module to extract channel context, and an identity path to propagate low-level spatial features. Afterwards, low-level spatial features and their spatial- and channel-wise contextualized representations are aggregated across a convolutional fusion module (CFM).}
 \label{fig:main_fig}
\end{figure*}

A recent alternative is transformer models that employ self-attention operators instead of convolution \cite{Wang_2023, yuan2023hcformer,9928347,JIANG2024106260}. Transformers process images as a sequence of tokens (i.e., image patches), and perform non-local filtering driven by inter-token similarities to improve sensitivity for long-range context. Note that evaluating similarity between all token pairs induces quadratic complexity with respect to sequence length, compromising computation and learning efficiency \cite{9940308}. While images can be downsampled or split into large-sized patches to reduce sequence length, this undesirably limits spatial precision \cite{9607618}. Common strategies to maintain a degree of local sensitivity in transformer-based methods have employed hybrid architectures that reserve high-resolution processing to convolutional branches \cite{JIANG2024106260, TransCT,yiyu2022low}, or architectures with locally-biased attention layers \cite{Wang_2023,jian2024swinct}. Since these approaches restrict the spatial resolution or range of attention operators, they typically suffer from a suboptimal trade-off between sensitivity to short- versus long-range context \cite{ssm_review1}. 

An emerging framework in machine learning that promises to efficiently capture long-range context while maintaining high local precision is based on state-space models (SSM) \cite{zhu2024vision}. SSMs process images as a sequence of pixels whose relationships are modeled recurrently under linear complexity with respect to sequence length, so they can in principle be an ideal candidate to process LDCT images \cite{vimednet}. However, conventional SSM modules adopted in previous imaging studies are devised to capture context exclusively across spatial dimensions \cite{liu2024vmamba}. Neglecting channel context in latent feature maps can cause poor use of interdependencies across feature channels, compromising quality of feature extraction and downstream task performance. Thus, existing SSM models can have limited utility in LDCT denoising, where sensitive capture of diverse contextual features in CT images is key to model performance. 

Here we introduce a novel SSM-based model, DenoMamba, to improve performance in LDCT image denoising by effectively capturing spatial and channel context in CT images without compromising local precision. To do this, DenoMamba leverages a novel architecture that cascades multiple FuseSSM blocks per network stage (Fig.~\ref{fig:main_fig}). The proposed FuseSSM blocks convolutionally fuse the spatial context captured by a spatial SSM module with the channel context captured by a novel channel SSM module (Fig.~\ref{fig:main_fig_b}). The proposed channel SSM module employs a secondary gated convolution network following the SSM layer in order to extract higher-order features of channel context. Meanwhile, to improve preservation of low-level spatial representations in LDCT images, FuseSSM blocks are equipped with an identity propagation path. These building blocks empower DenoMamba to capture diverse contextual information in LDCT images, without necessitating downsampling or patching procedures that restrict spatial precision in transformers. Comprehensive evaluations on LDCT datasets acquired at 25\% and 10\% of nominal radiation doses demonstrate the superior performance of DenoMamba compared to state-of-the-art baselines. Code to implement DenoMamba is publicly available at \href{https://github.com/icon-lab/DenoMamba}{https://github.com/icon-lab/DenoMamba}.

\vspace{4mm}
{\textbf{Contributions}}
\begin{itemize}
    \item To our knowledge, DenoMamba is the first LDCT denoising method that leverages state-space modeling across spatial and channel dimensions of latent feature maps. 
    \item DenoMamba employs a novel architecture based on convolutional fusion of feature maps extracted via spatial and channel SSM modules along with an identity propagation path, enabling it to effectively consolidate a comprehensive set of contextual features.  
    \item A novel channel SSM module is introduced that extracts higher-level features of channel context by cascading a transposed SSM layer operating over the channel dimension with a subsequent gated convolution network. 
\end{itemize}

\section{Related Work}

\subsection{Learning-based Models}
Earlier methods in LDCT denoising have adopted CNN models that process images via compact convolution operators \cite{8946589,2017arXiv170200288C,1704886}. The implicit locality bias of convolution enables CNNs to attain high computational efficiency, to learn effectively from modest size datasets, and to offer high sensitivity to short-range contextual features in medical images \cite{slater}. Yet, vanilla CNNs also manifest a number of key limitations; and a number of architectural improvements have been sought over the years to address them. To improve preservation of detailed tissue structure, models that separately process low- and high-frequency image components \cite{https://doi.org/10.1002/mp.12344,9669891}, models embodying Sobel convolutional layers to emphasize tissue boundaries \cite{9320928}, and multi-scale models that fuse features extracted at different scales have been proposed \cite{li2023multi}. To enhance denoising performance near rare pathology, multiplicative attention layers have been embedded in CNN models \cite{li2020}. In recent years, adversarial models \cite{8340157,wolterink2017tmi} and diffusion models \cite{IDDPM,gao2023corediff} based on CNN backbones have also been considered to further improve realism in denoised CT images by adopting generative learning procedures. While these advancements have helped push the performance envelope in LDCT denoising, CNN models often struggle to capture long-range contextual features in medical images due to the inherent locality bias of convolution operators \cite{kodali2018,pgan}. 

As an alternative to CNNs, recent studies have introduced transformer models that instead process images via non-local self-attention operators \cite{yuan2023hcformer,TransCT}. Driven by the similarities between all possible pairs of image tokens regardless of their spatial distance, self-attention operators enable transformers to offer exceptional sensitivity to long-range contextual features \cite{9940308,ssdiffrecon}. Yet, the quadratic complexity of self-attention with respect to image size has limited the spatial precision at which transformers can be utilized in practice. Aiming at this fundamental limitation, a group of studies have proposed hybrid approaches that alleviate model complexity by lowering feature map dimensions that are provided to the transformer modules \cite{10385173,yuan2023hcformer,TransCT,jian2024swinct} or by introducing loss terms to improve preservation of edge features \cite{JIANG2024106260,yiyu2022low}. That said, it remains a significant challenge in transformer-based methods to maintain a favorable balance between short- and long-range sensitivity in high-resolution medical images, without introducing heavy model complexity that can elevate computational burden and compromise learning efficacy \cite{ssm_review1}.

\begin{figure*}[!t]
\centerline{\includegraphics[width=0.77\textwidth]{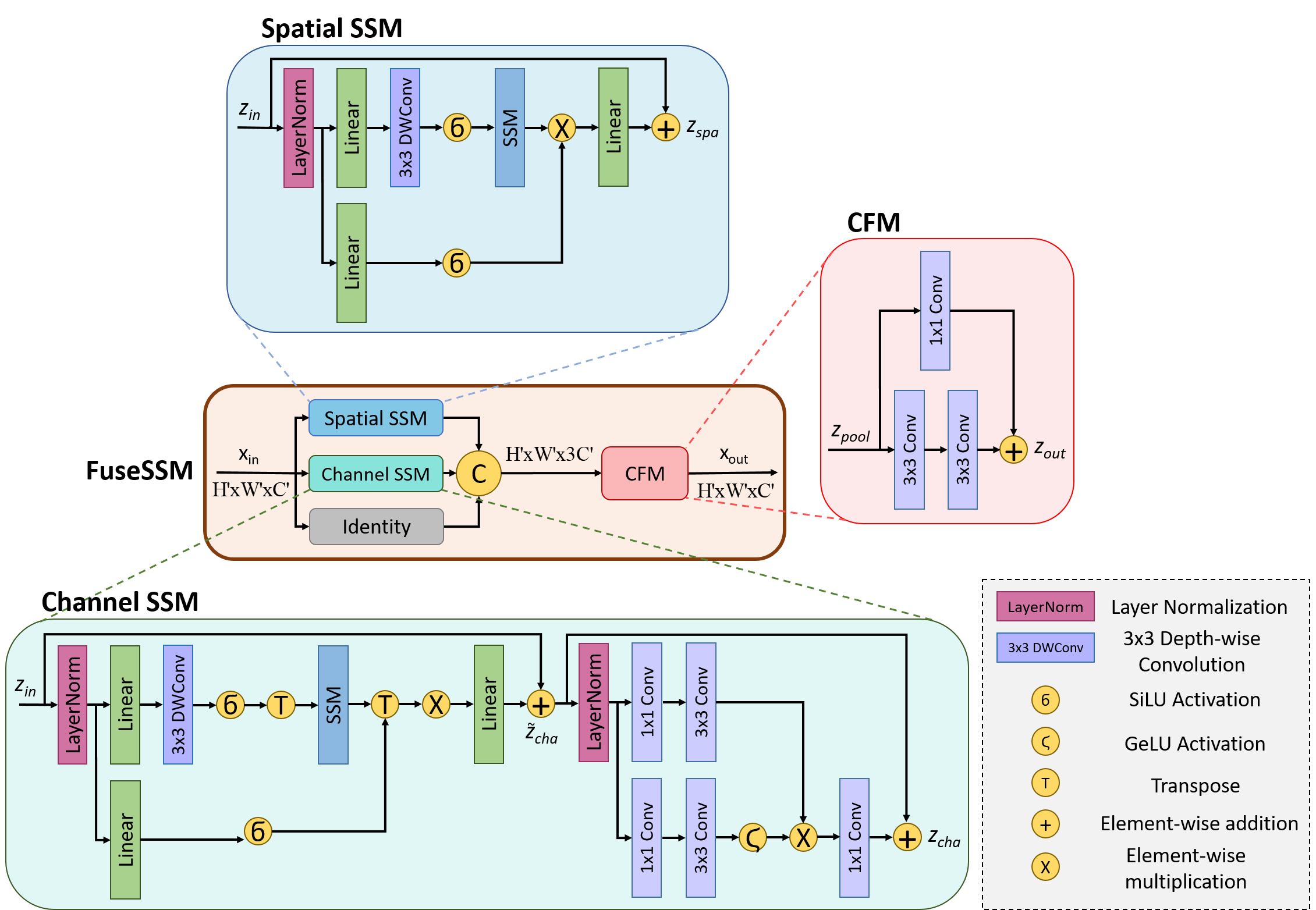}}
\caption{Inner modules of the FuseSSM blocks. Each FuseSSM block comprises a channel SSM module, a spatial SSM module, an identity propagation path, and a CFM module. The channel SSM module performs convolutional encoding of image tokens after layer normalization, and processes the transposed feature map via an SSM layer to capture an initial set of contextual features across the channel dimension. To further extract higher-order latent features, this initial set is projected through a gated convolutional network, and the two sets of contextual features are residually combined. The spatial SSM module performs convolutional encoding of image tokens after layer normalization, and processes the feature map via an SSM layer to capture contextual features across the spatial dimension. The CFM module pools low-level features propagated by the identity path with contextual features from the channel and spatial SSM modules, and nonlinearly fuses them via convolutional layers.}
 \label{fig:main_fig_b}
\end{figure*}
\subsection{SSM Models}
SSMs are an emerging framework in machine learning to efficiently capture long-range context without facing the significant complexity of attention operators, so they are less amenable to compromises in local precision \cite{ssm_review1}. Building on this framework, recent studies have devised SSM-based models for medical imaging tasks such as segmentation \cite{ma2024umamba,xing2024segmamba,liu2024swinumamba}, classification \cite{yue2024medmamba}, synthesis \cite{2024arXiv240514022A}, and reconstruction \cite{2024arXiv240218451H}. Although SSMs have shown promise in these challenging tasks, their application to medical image denoising remains relatively untapped, presenting a compelling avenue for further research. Accordingly, here we introduce DenoMamba as a novel SSM model for attaining improved performance in LDCT image denoising. With similar aims to DenoMamba, a recent imaging study has proposed a hybrid CNN-SSM model for LDCT denoising dubbed ViMEDNet \cite{vimednet}. Yet, DenoMamba carries key architectural differences that distinguish it from existing SSM-based methods. Specifically, ViMEDNet pools convolutional and SSM-based features of spatial context, and it uses conventional SSM modules that neglect channel context while processing feature maps \cite{vimednet}. In contrast, DenoMamba employs a novel architecture built exclusively on state-space operators, and it embodies dedicated spatial SSM and channel SSM modules that allow DenoMamba to capture both spatial and channel interdependencies efficiently within a unified framework. To our knowledge, DenoMamba is the first LDCT denoising method in the literature that leverages state-space modeling to simultaneously capture spatial and channel context in latent feature maps of CT images. Furthermore, DenoMamba employs novel channel SSM modules that capture higher-order feature of channel context via secondary gated convolutions subsequent to SSM layers. Collectively, these unique technical attributes enable DenoMamba to achieve high spatial precision while maintaining sensitivity to a diverse array of contextual features in LDCT images.

\section{Theory}

\subsection{Problem Definition}
LCDT image denoising involves suppression of elevated noise in low-dose CT scans due to reduced number of incident photons from the X-ray beam. Learning-based methods aim to solve this problem by training a neural network model to map noisy LDCT images onto denoised images that would be consistent with a normal dose CT (NDCT) scan. Let $x \in \mathbb{R}^{H \times W}$ denote the noisy LDCT image, and $y \in \mathbb{R}^{H \times W }$ denote the corresponding NDCT image, where \( H \), \( W \) are the image height and width, respectively. Given a training set of $T$ image pairs $( x_{tr}[i], y_{tr}[i] )$ with $i \in [1 \mbox{ } T]$, a network model \( f_{\theta}(\cdot) \) with parameters \( \theta \) can be trained as follows:
\begin{align}
    {\theta}^{*}=\text{argmin}_{\theta} \sum_{i=1}^{T}{\left\| f_{\theta}\left( x_{tr}[i] \right) - y_{tr}[i] \right\|^{2}_{2}}.
\end{align}
Upon successful training, the optimal parameters \( {\theta}^{*} \) that minimize the loss function should yield a model capable of effectively attenuating noise in LDCT images. The trained model can then be deployed to process novel LDCT images, generating denoised outputs as $\hat{y}_{test}[i] = f_{\theta^{*}}\left( x_{test}[i] \right)$.

\subsection{DenoMamba}
DenoMamba is the first LDCT image denoising method in the literature that uses SSMs to model spatial and channel context, to our knowledge. It employs a novel architecture based on FuseSSM blocks that aggregate low-level spatial features along with a comprehensive set of contextual features across spatial and channel dimensions, hence maintaining a favorable balance between short- and long-range sensitivity. In the following subsections, we describe the overall architecture of DenoMamba and the inner structure of FuseSSM blocks.

\subsubsection{Overall Model Architecture} As depicted in Fig. \ref{fig:main_fig}, DenoMamba follows an hourglass structure with $K$ encoder and $K$ decoder stages. Each stage is implemented as a cascade of multiple FuseSSM blocks. Starting from the noisy LDCT image $x$ taken as model input, encoder stages serve to extract latent contextualized representations via FuseSSM blocks and to resample the feature map dimensions. Let $x_{\text{enc}}^k$ denote the feature map at the output of the $k$th encoder stage, with $k \in [ 1, 2, ..., K]$ and $x_{\text{enc}}^0 = x$. The mapping through the $k$th encoder stage can be described as follows:
\begin{equation}
x_{\text{enc}}^{k} = \begin{cases}
& \text{Down} (\text{Enc}^{k}\left( x_{\text{enc}}^{k-1};\theta^{k}_{enc} \right)),  \text{ if } k \neq K  \\  
& \text{Enc}^{k}\left( x_{\text{enc}}^{k-1};\theta^{k}_{enc} \right), \text{ if } k = K
\end{cases}
\end{equation}
where $\text{Enc}^k (\cdot) := \bigoplus_{r=1}^{E(k)} \mathrm{FuseSSM} (\cdot)$ denotes composition of the $k$th stage via recursive application of $E(k)$ FuseSSM blocks, $\theta^{k}_{enc}$ denotes the parameters of these FuseSSM blocks, $\text{Down}(\cdot)$ denotes a learnable downsampling operator, and $x_{\text{enc}}^{k} \in \mathbb{R}^{\frac{H}{2^k} \times \frac{W}{2^k} \times 2^k C}$. Note that downsampling is performed at all encoder stages, except for the final stage (i.e., $k=K$). 

Starting from the encoded feature map $x_{\text{enc}}^{K}$, decoder stages then serve to recover a denoised image $\hat{y}$ from the latent representations via a cascade of FuseSSM blocks and resampling of feature map dimensions. The decoder stages follow a mirror-reversed order, such that $x_{\text{dec}}^k$ denotes the feature map at the output of the $k$th decoder stage, with $k \in [ K, K-1, ..., 1]$ and $x_{\text{dec}}^K = x_{\text{enc}}^K$. Thus, the mapping through the $kth$ decoder stage can be described as follows: 
\begin{equation}
x_{\text{dec}}^{k-1} = \begin{cases}
& \text{Dec}^{k}\left( \text{Up}(x_{\text{dec}}^{k}) + x_{\text{enc}}^{k-1} ;\theta^{k}_{dec} \right),  \text{ if } k \neq 1  \\  
& \text{Dec}^{k}\left( x_{\text{dec}}^{k} + x_{\text{enc}}^{k-1}; \theta^{k}_{dec} \right), \text{ if } k = 1
\end{cases}
\end{equation}
where $\text{Dec}^k (\cdot) := \bigoplus_{r=1}^{D(k)} \mathrm{FuseSSM} (\cdot)$ denotes composition of the $k$th stage via recursive application of $D(k)$ FuseSSM blocks, $\theta^{k}_{dec}$ denotes the parameters of FuseSSM blocks in the $k$th decoder stage, $\text{Up}(\cdot)$ denotes a learnable upsampling operator, and $x_{\text{dec}}^{k-1} \in \mathbb{R}^{\frac{H}{2^{k-2}} \times \frac{W}{2^{k-2}} \times 2^{k-2} C}$. Note that upsampling is performed on the decoder feature map $x_{\text{dec}}^{k}$ in the beginning of all decoder stages, except for the final stage (i.e., $k=1$). Furthermore, encoder feature maps from the respective encoder stage $x_{\text{enc}}^{k-1}$ are residually added onto the input decoder maps to improve preservation of low-level structural representations in LDCT images. The final output of DenoMamba is taken as $\hat{y} = x_{\text{dec}}^{0}$.  

\subsubsection{FuseSSM blocks} DenoMamba is constructed with novel FuseSSM blocks that comprise a spatial SSM module to capture contextual representations in the spatial domain and a channel SSM module to capture contextual representations in the channel domain \cite{liu2024vmamba}. We propose to project input feature maps across three parallel pathways that propagate the contextualized representations from spatial and channel SSM modules, along with original input features. Afterwards, these representations are merged via a convolutional fusion module (CFM). For a given FuseSSM block, a schematic of the individual components are depicted in Fig. \ref{fig:main_fig_b}.

The design of FuseSSM blocks in encoder and decoder stages are identical apart from variability in feature map dimensions. Thus, here we will describe the projections through a FuseSSM block without distinguishing between encoder/decoder stages. Assuming that the input feature map at the $k$th stage is $z_{in}=x^{k} \in \mathbb{R}^{H' \times W' \times C'}$, the respective FuseSSM block first projects the input through three parallel pathways to compute contextualized representations: 
\begin{equation}
    \{ z_{\text{spa}}, z_{\text{cha}}, z_{\text{in}} \} = \{ \text{SSM}_{\text{spa}}( z_{\text{in}}),  \text{SSM}_{\text{cha}} ( z_{\text{in}} ) , \text{I} ( z_{\text{in}} )\},
\end{equation}
where \( \text{SSM}_{\text{spa}} \) denotes the spatial SSM, \( \text{SSM}_{\text{cha}} \) denotes the channel SSM, and I denotes the identity propagation path. The extracted contextual representations are then pooled and convolutionally fused within the CFM module: 
\begin{align}
    z_{\text{pool}} &= \text{Concat} (z_{\text{spa}}, z_{\text{cha}}, z_{\text{in}}),\\
    z_{\text{out}} &=  \text{Conv}^{1\times 1}\left( z_{\text{pool}} \right) \oplus \text{Conv}^{3\times 3}\left( \text{Conv}^{3\times 3}\left( z_{\text{pool}} \right) \right),
\end{align}
where Concat denotes a concatenation operator that pools feature maps across the channel dimension, $\text{Conv}^{1 \times 1}$ and $\text{Conv}^{3 \times 3}$ respectively denote $1 \times 1$ and $3 \times 3$ convolutional layers, and $\oplus$ is the element-wise addition operator. The feature map $z_{\text{out}} \in \mathbb{R}^{H' \times W' \times C'}$ is taken as the output of the FuseSSM block. 

\textbf{Spatial SSM:} Within the spatial SSM module, a first branch linearly embeds the input map and uses a nonlinearity to produce a gating variable \( GP_{\text{spa}} \in \mathbb{R}^{H' \times W' \times \alpha C'}\):
\begin{align}
    GP_{\text{spa}} = \sigma(f_{\text{lin}}(z_{\text{in}})),
\end{align}
where $\sigma$ is a SiLU activation function and \( f_{\text{lin}}\) denotes a learnable linear mapping that expands the feature map across the channel dimension by a factor $\alpha$. A second branch performs linear embedding and convolutional encoding, followed by an SSM layer to derive \(M_{\text{spa}}\in \mathbb{R}^{H' \times W' \times \alpha C'}\): 
\begin{align}
    M_{\text{spa}} = \text{SSM}\left(\sigma(\text{DWConv}^{3\times3}(f_{\text{lin}}(z_{\text{in}})))\right),
\end{align}
where $\text{SSM}$ denotes a state-space layer, \( \text{DWConv}^{3\times3} \) refers to depth-wise convolution of kernel size \( 3\times 3 \). 

Here, the state-space layer is implemented based on the Mamba variant in \cite{liu2024vmamba}. Accordingly, scanning is performed across two spatial dimensions of the input feature map to the SSM layer in order to expand it onto a sequence $s \in \mathbb{R}^{H'W' \times \alpha C'}$. The sequence is then processed via a discrete state-space model independently for each channel:
\begin{align}
    h[n] &= \mathbf{A} h[n-1] + \mathbf{B} s[n], \\
    \bar{s}[n] &= \mathbf{C} h[n],
\end{align}
where $n \in [1 \mbox{ } H'W']$ is an integer denoting sequence index, \(h\) denotes the hidden state, \(s[n]\) is the $n$th element of the input sequence. \(\mathbf{A} \in \mathbb{R}^{N,N}\), \(\mathbf{B} \in \mathbb{R}^{N,1}\), \(\mathbf{C} \in \mathbb{R}^{1,N}\) are learnable parameters of the state-space model, with \(N\) indicating the hidden dimensionality. Note that \(\mathbf{B}\) and \(\mathbf{C}\) are taken to be functions of the input sequence in Mamba to enable input-adaptive processing \cite{liu2024vmamba}. The output sequence  \(\bar{s} \in \mathbb{R}^{H'W' \times \alpha C'}\) is remapped back onto the feature map \(M_{\text{spa}}\in \mathbb{R}^{H' \times W' \times \alpha C'}\).

To compute the module output, \( M_{\text{spa}} \) is gated with $GP_{\text{spa}}$, and the result is linearly projected and combined with the input through a residual connection:
\begin{align}
    z_{\text{spa}} = z_{\text{in}}+f_{\text{lin}}(GP_{\text{spa}} \odot M_{\text{spa}}),
\end{align}
where $\odot$ denotes the Hadamard product operator, and $f_{\text{lin}}$ is devised to use an expansion factor of $1/\alpha$ such that $z_{\text{spa}}\in \mathbb{R}^{H' \times W' \times C'}$ has matching dimensionality to $z_{\text{in}}$.

\textbf{Channel SSM:} Similar to the spatial SSM module, within the channel SSM module, a first branch produces a gating variable and a second branch performs state-space modeling on the sequentialized input feature map to capture contextual interactions in the channel dimension:
\begin{align}
    GP_{\text{cha}} &= \sigma(f_{\text{lin}}(z_{\text{in}})),\\ 
    M_{\text{cha}} &=\text{SSM}\left( (\sigma(\text{DWConv}^{3\times3}(f_{\text{lin}}(z_{\text{in}}))))^\top \right)^\top,
\end{align}
where $\top$ denotes the transpose operator. Differing from the spatial SSM module, the channel SSM module captures channel context by transposing the input sequence prior to and after the SSM layer. This results in an intermediate set of contextual representations $\Tilde{z}_{\text{cha}}\in \mathbb{R}^{H' \times W' \times C'}$ derived as:
\begin{equation}
       \Tilde{z}_{\text{cha}} = z_{\text{in}}+f_{\text{lin}}(GP_{\text{cha}} \odot M_{\text{cha}}). 
\end{equation}
Note that many layers in DenoMamba can perform spatial encoding, such as the depth-wise convolutional layers in FuseSSM blocks and downsampling/upsampling layers across encoder/decoder stages. Collectively, these layers can learn a hierarchy of latent features of spatial context. Yet, channel encoding is primarily performed in the SSM layers of the channel SSM module, limiting the information captured on channel context. To address this limitation, here we propose a novel channel SSM module that incorporates a gated convolution network to extract latent features of channel context. For this purpose, a second gating variable $GP^2_{\text{cha}}\in \mathbb{R}^{H' \times W' \times C'}$ is first computed:
\begin{align}
GP^2_{\text{cha}} = \zeta(\text{DWConv}^{3\times 3}(\text{Conv}^{1\times 1}(\Tilde{z}_{\text{cha}})),
\end{align}
where $\zeta$ is an ReLU activation function. $GP^2_{\text{cha}}$ is then used to modulate latent features of $\Tilde{z}_{\text{cha}}$:
\begin{equation}
  z_{\text{cha}} = \text{Conv}^{1\times 1}( GP^2_{\text{cha}} \odot \text{DWConv}^{3\times 3}(\text{Conv}^{1\times 1}(\Tilde{z}_{\text{cha}})))) + \Tilde{z}_{\text{cha}}.
\end{equation}
As such, the module output $z_{\text{cha}}\in \mathbb{R}^{H' \times W' \times C'}$ has matching dimensionality to $z_{\text{in}}$.

\subsubsection{Learning Procedures} 
Given a training set of image pairs $( x_{tr}[i], y_{tr}[i] )$ with $i \in [1 \mbox{ } T]$, DenoMamba with parameters \( \theta_{\text{enc}},\theta_{\text{dec}} \) is trained via a pixel-wise $\ell_1$-loss term:
\begin{multline}
    \{\theta_{\text{enc}}^{*},\theta_{\text{dec}}^{*}\} = \text{argmin}_{\theta_{\text{enc}},\theta_{\text{dec}}} \sum_{i=1}^{T} \bigg\| \text{Dec}^{(K:1)} \Big( \\ 
    \text{Enc}^{(1:K)} \big(
    x_{tr}[i]; \theta_{\text{enc}}^{(1:K)} \big); \theta_{\text{dec}}^{(K:1)} \Big) - y_{tr}[i] \bigg\|_{1}.
\end{multline}
Using the trained parameters \(  \{\theta_{\text{enc}}^{*},\theta_{\text{dec}}^{*}\} \), the model can be deployed to process a novel LDCT image from the test set $x_{test}[i]$ to estimate a denoised output $\hat{y}_{test}[i]$ as:
\begin{equation}
\hat{y}_{test}[i] = \text{Dec}^{(K:1)} \Big( \text{Enc}^{(1:K)} \big( \\ x_{test}[i]; {\theta}_{\text{enc}}^{* \, (1:K)} \big); {\theta}_{\text{dec}}^{* \, (K:1)} \Big)
\end{equation}

\section{Experimental Setup}

\subsection{Datasets}
\textit{AAPM Dataset:} Demonstrations of denoising performance were conducted on contrast-enhanced abdominal CT scans from the 2016 AAPM-NIBIB-MayoClinic Low Dose CT Grand Challenge \cite{mccollough2016tu}. Two different dose reduction levels were considered, resulting in 25\%- and 10\%-dose datasets. Normal dose CT (NDCT) scans were acquired at 120 kV reference tube potential with 200 effective mAs as quality reference. LDCT at 25\%-dose with 50 effective mAs and LDCT at 10\%-dose with 20 effective mAs were simulated from NDCT images assuming a Poisson-Gaussian noise distribution \cite{9940308,10472025}. The training set comprised 760 NDCT-LDCT image pairs, the validation set had 35 pairs, and the test set had 200 pairs. There was no subject overlap among the three sets, and each set contained a mixture of CT images reconstructed at either 1 mm or 3 mm slice thickness. All images were resized to 256$\times$256 in-plane resolution. 

\textit{Piglet CT Dataset:} This dataset contained CT scans of a deceased piglet acquired at varying radiation doses attained by adjusting the tube current \cite{pigletdata}. NDCT scans were acquired at 100 kV reference tube potential with 300 effective mAs as quality reference radiation dose. LDCT scans were acquired at 10\%-dose by prescribing 30 effective mAs. As this dataset was primarily used for evaluating the generalization performance of models trained on the AAPM dataset, we only curated a test set comprising 350 NDCT-LDCT image pairs. All images had 0.625 mm slice thickness, and they were resized to 256$\times$256 in-plane resolution.

\begin{figure*}[t]
\centerline{\includegraphics[width=0.88\textwidth]{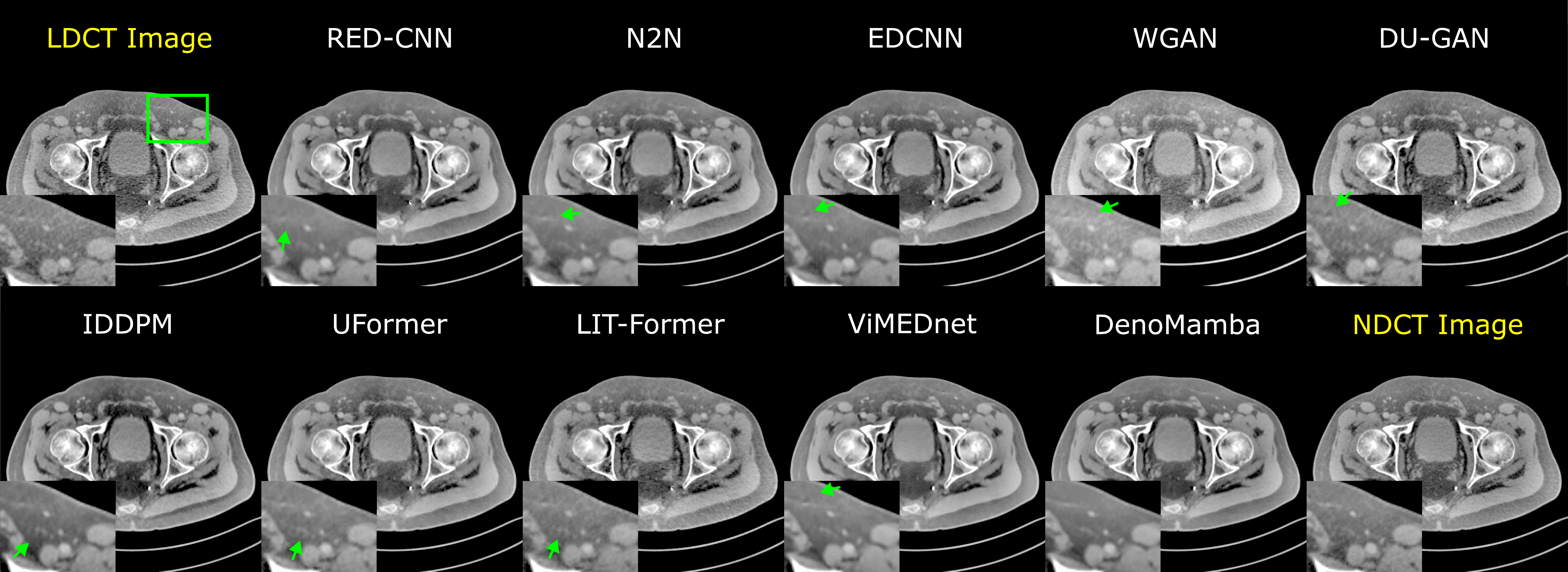}}
\caption{Denoising results from the 25\%-dose AAPM dataset are depicted for representative cross-sections. Images recovered by competing methods are shown along with the LDCT image (i.e., model input), and the NDCT image (i.e., ground truth). Zoom-in displays and arrows are used to showcase regions with visible differences in image quality among competing methods. Display windows of [-150 350] HU are used.}
\label{fig:ldct_1}
\end{figure*}

\begin{table}[t]
\centering
\setlength\tabcolsep{3.5pt}
\renewcommand{\arraystretch}{1.4}
\caption{Denoising performance of competing methods on the 25\%-dose AAPM dataset. PSNR (dB), SSIM (\%) and RMSE (\%) metrics are listed as mean$\pm$std across the test sets. Boldface marks the method that offers the best performance for each metric.}
\resizebox{0.875\columnwidth}{!}{%
\begin{tabular}{|c|ccc|}
\hline
\multirow{2}{*}{}                & \multicolumn{1}{c|}{$\uparrow$ \textbf{PSNR (dB)}}  & \multicolumn{1}{c|}{$\uparrow$ \textbf{SSIM (\%)}}  & $\downarrow$ \textbf{RMSE (\%)}    \\ \hline
\multicolumn{1}{|c|}{RED-CNN}    & \multicolumn{1}{c|}{41.02 ± 3.03} & \multicolumn{1}{c|}{96.25 ± 1.65} & 9.54 ± 0.32  \\ \hline
\multicolumn{1}{|c|}{N2N}        & \multicolumn{1}{c|}{40.72 ± 2.98} & \multicolumn{1}{c|}{96.37 ± 1.67} & 9.85 ± 0.32  \\ \hline
\multicolumn{1}{|c|}{EDCNN}      & \multicolumn{1}{c|}{40.86 ± 3.06} & \multicolumn{1}{c|}{96.07 ± 1.72} & 9.67 ± 0.32  \\ \hline
\multicolumn{1}{|c|}{WGAN}       & \multicolumn{1}{c|}{39.79 ± 2.54} & \multicolumn{1}{c|}{94.80 ± 2.29} & 10.59 ± 0.35 \\ \hline
\multicolumn{1}{|c|}{DU-GAN}     & \multicolumn{1}{c|}{40.01 ± 3.11} & \multicolumn{1}{c|}{94.48 ± 3.13} & 10.92 ± 0.41 \\ \hline
\multicolumn{1}{|c|}{IDDPM}      & \multicolumn{1}{c|}{41.04 ± 2.22} & \multicolumn{1}{c|}{96.55 ± 1.66} & 9.40 ± 0.29  \\ \hline
\multicolumn{1}{|c|}{UFormer}    & \multicolumn{1}{c|}{41.05 ± 2.79} & \multicolumn{1}{c|}{96.76 ± 1.64} & 9.43 ± 0.33  \\ \hline
\multicolumn{1}{|c|}{LIT-Former} & \multicolumn{1}{c|}{40.93 ± 2.82} & \multicolumn{1}{c|}{96.05 ± 1.87} & 9.62 ± 0.33  \\ \hline
\multicolumn{1}{|c|}{ViMEDnet}   & \multicolumn{1}{c|}{41.73 ± 3.12} & \multicolumn{1}{c|}{96.24 ± 1.68} & 8.86 ± 0.36  \\ \hline
\multicolumn{1}{|c|}{\textbf{DenoMamba}} & \multicolumn{1}{c|}{\textbf{42.69 ± 2.85}} & \multicolumn{1}{c|}{\textbf{97.07 ± 1.74}} & \textbf{8.00 ± 0.33}  \\ \hline
\end{tabular}%
}
\label{tab:main1}
\end{table}

\subsection{Architectural Details}
In DenoMamba, a $K=4$ stage encoder-decoder architecture was used, where the number of FuseSSM blocks cascaded within a given stage varied as $E$ = [4, 6, 6, 8] across encoder stages and as $D$ = [6, 6, 4, 2] across decoder stages, respectively. Spatial resolution was lowered by a factor of 2 in each encoder stage except for the final one, while the channel dimensionality was set as [48, 96, 192, 384] across stages. Conversely, spatial resolution was increased by a factor of 2 in each decoder stage except for the final one, with the channel dimensionality set as [192, 96, 48, 48] across stages. Both spatial and channel SSM modules used a state expansion factor of $N$=16, a local convolution width of 4, and a block expansion factor of $\alpha$=2.  

\subsection{Competing Methods}
\label{sec:methods}
We demonstrated DenoMamba against several state-of-the-art methods for LDCT denoising. For fair comparisons, all competing methods were implemented with a pixel-wise $\ell_1$-loss similar to DenoMamba. The only exceptions to this were generative models that were implemented with their original loss terms required to enable adversarial or diffusive learning. 

\textit{RED-CNN}: A convolutional model was considered that uses a hierarchical encoder-decoder architecture equipped with shortcut connections \cite{2017arXiv170200288C}. 

\textit{N2N}: A convolutional model was considered that was originally proposed for self-supervised learning on noisy CT images \cite{9577596}. For fair comparison, the architecture of N2N was adopted to perform supervised learning.   

\textit{EDCNN}: A convolutional model was considered that employs a trainable Sobel convolution kernel for edge detection and dense connections \cite{9320928}.

\textit{WGAN}: An adversarial model that uses convolutional generator and discriminator subnetworks was considered \cite{8340157}. Loss term weights were set as $\lambda$ = 10, $\lambda_1$ = 0.1, $\lambda_2$ = 0.1.

\textit{DU-GAN}: An adversarial model that uses convolutional generator and discriminator subnetworks was considered \cite{9617604}. Loss terms weights were set as $\lambda_{adv}$ = 0.1, $\lambda_{img}$ = 1, and $\lambda_{grd}$ = 20.

\textit{IDDPM}: A diffusion model with a convolutional backbone augmented with attention mechanism was considered that generated NDCT images starting from Gaussian noise images, with additional guidance from the LCDT image provided as input \cite{IDDPM}. The number of diffusion steps was taken as 1000. 

\textit{UFormer}: An efficient transformer model was considered that uses a hierarchical encoder-decoder architecture and local window-based self-attention \cite{9878729}. 

\textit{LIT-Former}: An efficient transformer model was considered that was originally proposed for processing 3D images with separate transformer modules for in-plane and through-plane dimensions \cite{10385173}. LIT-former was adopted for 2D images by removing the through-plane modules. 

\textit{ViMEDNet}: A state-space model was considered that uses a hierarchical encoder-decoder architecture equipped with spatial SSM modules \cite{vimednet}.

\begin{figure*}[t]
\centerline{\includegraphics[width=0.88\textwidth]{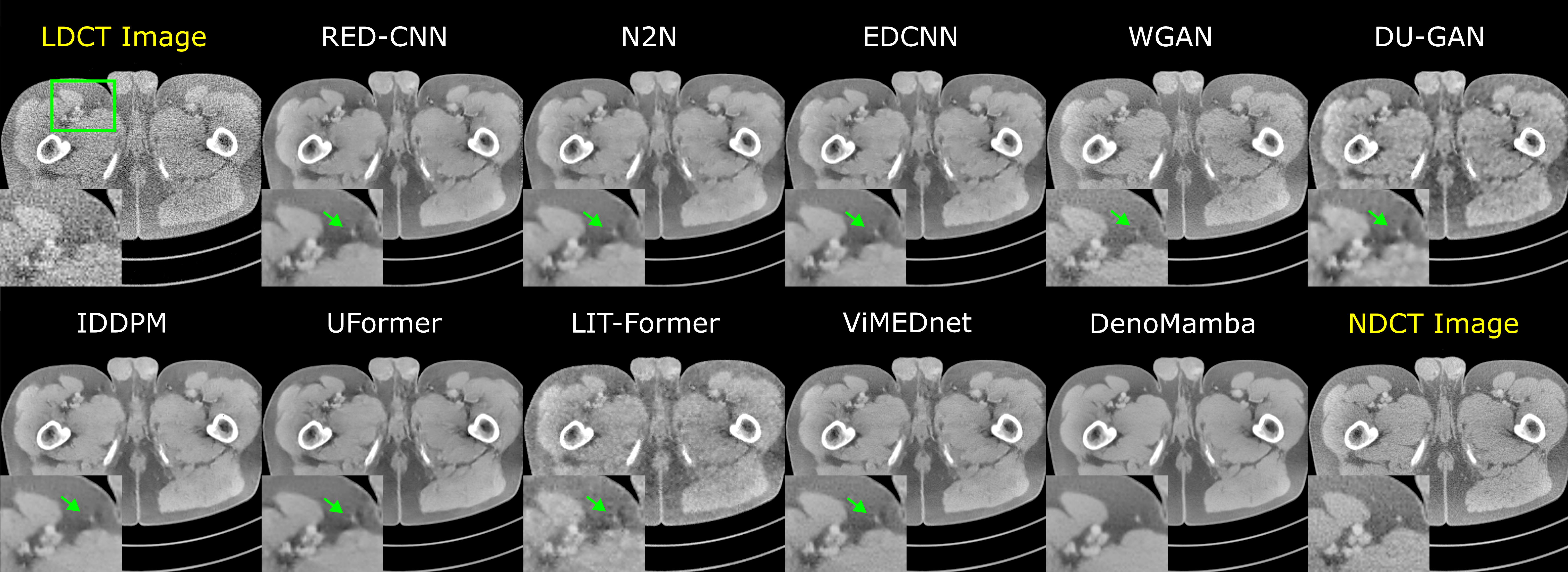}}
\caption{Denoising results from the 10\%-dose AAPM dataset are depicted for representative cross-sections. Display windows of [-350 350] HU are used.}
\label{fig:ldct_2}
\end{figure*}
\begin{table}[t]
\centering
\setlength\tabcolsep{3.5pt}
\renewcommand{\arraystretch}{1.4}
\caption{Denoising performance of competing methods on the 10\%-dose AAPM dataset.}
\resizebox{0.875\columnwidth}{!}{%
\begin{tabular}{|c|ccc|}
\hline
\multirow{2}{*}{}                & \multicolumn{1}{c|}{$\uparrow$ \textbf{PSNR (dB)}}  & \multicolumn{1}{c|}{$\uparrow$ \textbf{SSIM (\%)}}  & $\downarrow$ \textbf{RMSE (\%)}    \\ \hline
\multicolumn{1}{|c|}{RED-CNN}    & \multicolumn{1}{c|}{38.27 ± 2.39} & \multicolumn{1}{c|}{95.18 ± 1.65} & 12.79 ± 0.35  \\ \hline
\multicolumn{1}{|c|}{N2N}        & \multicolumn{1}{c|}{37.52 ± 2.41} & \multicolumn{1}{c|}{94.74 ± 1.74} & 13.75 ± 0.35  \\ \hline
\multicolumn{1}{|c|}{EDCNN}      & \multicolumn{1}{c|}{37.80 ± 2.49} & \multicolumn{1}{c|}{94.10 ± 1.78} & 13.39 ± 0.37  \\ \hline
\multicolumn{1}{|c|}{WGAN}       & \multicolumn{1}{c|}{37.37 ± 2.19} & \multicolumn{1}{c|}{94.22 ± 1.97} & 13.70 ± 0.38  \\ \hline
\multicolumn{1}{|c|}{DU-GAN}     & \multicolumn{1}{c|}{37.57 ± 2.46} & \multicolumn{1}{c|}{94.24 ± 2.88} & 13.91 ± 0.47  \\ \hline
\multicolumn{1}{|c|}{IDDPM}      & \multicolumn{1}{c|}{38.16 ± 2.60} & \multicolumn{1}{c|}{94.88 ± 1.73} & 12.79 ± 0.41  \\ \hline
\multicolumn{1}{|c|}{UFormer}    & \multicolumn{1}{c|}{38.77 ± 2.62} & \multicolumn{1}{c|}{95.82 ± 1.62} & 12.08 ± 0.41  \\ \hline
\multicolumn{1}{|c|}{LIT-Former} & \multicolumn{1}{c|}{37.33 ± 1.97} & \multicolumn{1}{c|}{92.47 ± 1.50} & 13.89 ± 0.34  \\ \hline
\multicolumn{1}{|c|}{ViMEDnet}   & \multicolumn{1}{c|}{38.88 ± 2.44} & \multicolumn{1}{c|}{95.90 ± 1.72} & 12.00 ± 0.39  \\ \hline
\multicolumn{1}{|c|}{\textbf{DenoMamba}} & \multicolumn{1}{c|}{\textbf{39.72 ± 2.43}} & \multicolumn{1}{c|}{\textbf{96.24 ± 1.73}} & \textbf{10.86 ± 0.38}  \\ \hline
\end{tabular}%
}
\label{tab:main2}
\end{table}

\subsection{Modeling Procedures}
Models were implemented using the PyTorch framework and trained on an NVidia RTX 3090 GPU. Training was performed via the Adam optimizer with parameters \(\beta_1=0.5\) and \(\beta_2=0.999\) \cite{adam}.  For all competing methods, the learning rate was set to $1 \times 10^{-4}$, and the number of epochs was set to 100. The initial learning rate was halved after every 30 epochs to promote gradual model refinement. Data were split into training, validation and test sets with no subject-level overlap between the three sets. Key model hyperparameters were selected via cross-validation for each competing method. Model performance was then evaluated on the test set with Peak Signal-to-Noise Ratio (PSNR), Structural Similarity Index (SSIM), and Root Mean Square Error (RMSE) metrics. Note that higher values of PSNR and SSIM, and lower values of RMSE indicate improved model performance. Significance of differences between competing methods were evaluated via non-parametric Wilcoxon signed-rank tests (p$<$0.05).

\section{Results}

\subsection{Comparison Studies}
\label{sec:comparison}
We demonstrated DenoMamba on abdominal CT scans from the  2016 AAPM Low Dose CT Grand Challenge via comparisons against several state-of-the-art methods from the LDCT denoising literature. Specifically, convolutional models (RED-CNN, N2N, EDCNN), generative models based on adversarial or diffusion learning (WGAN, DU-GAN, IDDPM), and contextually-sensitive models with efficient transformer or SSM backbones (UFormer, LIT-Former, ViMEDNet) were considered. While this study primarily focuses on the utility of network architectures for LDCT denoising, generative models were included in comparisons for a more comprehensive assessment (see Sec.~\ref{sec:methods} for further details on competing methods). Experiments were first conducted on the 25\%-dose dataset to recover NDCT images from LDCT measurements. Table \ref{tab:main1} lists performance metrics for competing methods on the test set. We find that DenoMamba significantly outperforms each competing method (p$<$0.05). On average, DenoMamba achieves performance improvements of 1.8dB PSNR, 0.8\% SSIM, 1.7\% RMSE over convolutional baselines; 2.4dB PSNR, 1.8\% SSIM, 2.3\% RMSE over generative baselines, and 1.5dB PSNR, 0.7\% SSIM, 1.3\% RMSE over contextually-sensitive baselines.

\begin{table*}[t]
\centering
\caption{Denoising performance of competing methods on the 10\%-dose Piglet CT dataset. Models trained on either the 25\%-dose (left panel) or 10\%-dose (right panel) AAPM scans were evaluated on Piglet CT scans.}
\resizebox{1.5\columnwidth}{!}{%
\begin{tabular}{c|ccc|ccc|}
\cline{2-7}
\multirow{2}{*}{}                & \multicolumn{3}{c|}{25\%-dose AAPM $\rightarrow$ 10\%-dose Piglet}  & \multicolumn{3}{c|}{10\%-dose AAPM $\rightarrow$ 10\%-dose Piglet}
\\ \cline{2-7} 
                                 & \multicolumn{1}{c|}{ $\uparrow$ \textbf{PSNR (dB)}}  & \multicolumn{1}{c|}{ $\uparrow$ \textbf{SSIM (\%)}}   & $\downarrow$ \textbf{RMSE (\%)}   & \multicolumn{1}{c|}{ $\uparrow$ \textbf{PSNR (dB)}}  & \multicolumn{1}{c|}{ $\uparrow$ \textbf{SSIM (\%)}}   & $\downarrow$ \textbf{RMSE (\%)}   \\ \hline
\multicolumn{1}{|c|}{RED-CNN}  & \multicolumn{1}{c|}{38.37 ± 3.96} & \multicolumn{1}{c|}{96.65 ± 3.26} & 32.76 ± 1.12  & \multicolumn{1}{c|}{38.66 ± 3.91} & \multicolumn{1}{c|}{96.42 ± 3.44} & 31.20 ± 1.21  \\ \hline
\multicolumn{1}{|c|}{N2N}      & \multicolumn{1}{c|}{37.69 ± 3.40} & \multicolumn{1}{c|}{95.86 ± 3.01} & 36.74 ± 1.16  & \multicolumn{1}{c|}{38.19 ± 3.46} & \multicolumn{1}{c|}{96.03 ± 3.13} & 32.61 ± 1.28 \\ \hline
\multicolumn{1}{|c|}{EDCNN}  & \multicolumn{1}{c|}{38.93 ± 4.07} & \multicolumn{1}{c|}{96.91 ± 3.07} & 29.93 ± 1.21     & \multicolumn{1}{c|}{38.78 ± 3.84} & \multicolumn{1}{c|}{96.88 ± 3.26} & 30.92 ± 1.24 \\ \hline
\multicolumn{1}{|c|}{WGAN}    & \multicolumn{1}{c|}{34.36 ± 3.07} & \multicolumn{1}{c|}{88.25 ± 3.69} & 47.11 ± 1.33   & \multicolumn{1}{c|}{35.11 ± 2.91} & \multicolumn{1}{c|}{88.40 ± 3.38} & 47.11 ± 1.35  \\ \hline
\multicolumn{1}{|c|}{DU-GAN}   & \multicolumn{1}{c|}{38.42 ± 3.49} & \multicolumn{1}{c|}{96.03 ± 5.22} & 31.28 ± 1.64  & \multicolumn{1}{c|}{38.34 ± 3.43} & \multicolumn{1}{c|}{96.29 ± 4.87} & 32.05 ± 1.48  \\ \hline
\multicolumn{1}{|c|}{IDDPM}    & \multicolumn{1}{c|}{38.58 ± 2.95} & \multicolumn{1}{c|}{96.52 ± 2.68} & 30.89 ± 0.86  & \multicolumn{1}{c|}{38.73 ± 2.73} & \multicolumn{1}{c|}{97.38 ± 2.59} & 30.82 ± 1.03  \\ \hline
\multicolumn{1}{|c|}{UFormer}   & \multicolumn{1}{c|}{38.44 ± 3.72} & \multicolumn{1}{c|}{96.70 ± 2.71} & 32.11 ± 1.11  & \multicolumn{1}{c|}{38.45 ± 3.45} & \multicolumn{1}{c|}{95.99 ± 2.74} & 31.72 ± 1.08 \\ \hline
\multicolumn{1}{|c|}{LIT-Former} & \multicolumn{1}{c|}{38.69 ± 3.11} & \multicolumn{1}{c|}{96.63 ± 3.10} & 30.17 ± 0.92 & \multicolumn{1}{c|}{38.53 ± 3.24} & \multicolumn{1}{c|}{96.50 ± 2.90} & 31.56 ± 0.97 \\ \hline
\multicolumn{1}{|c|}{ViMEDnet}  & \multicolumn{1}{c|}{39.05 ± 3.86} & \multicolumn{1}{c|}{97.31 ± 2.88} & 29.55 ± 1.17 & \multicolumn{1}{c|}{39.08 ± 3.83} & \multicolumn{1}{c|}{97.51 ± 2.76} & 29.68 ± 1.15  \\ \hline
\multicolumn{1}{|c|}{\textbf{DenoMamba}} & \multicolumn{1}{c|}{\textbf{39.88 ± 3.73}} & \multicolumn{1}{c|}{\textbf{98.40 ± 2.92}} & \textbf{28.82 ± 1.10} & \multicolumn{1}{c|}{\textbf{39.51 ± 3.53}} & \multicolumn{1}{c|}{\textbf{98.19 ± 2.81}} & \textbf{28.70 ± 1.12}  \\ \hline
\end{tabular}%
}
\label{tab:gen1}
\end{table*}

\begin{table*}[t]
\centering
\caption{Denoising performance of competing methods on the AAPM dataset. Models trained at 25\%-dose were tested at 10\%-dose (i.e., 25\%-dose $\rightarrow$ 10\%-dose), and models trained at 10\%-dose were tested at 25\%-dose (i.e., 10\%-dose $\rightarrow$ 25\%-dose).}
\resizebox{1.5\columnwidth}{!}{%
\begin{tabular}{c|ccc|ccc|}
\cline{2-7}
\multirow{2}{*}{}                & \multicolumn{3}{c|}{25\%-dose $\rightarrow$ 10\%-dose}                               & \multicolumn{3}{c|}{10\%-dose $\rightarrow$ 25\%-dose}
\\ \cline{2-7} 
                                 & \multicolumn{1}{c|}{ $\uparrow$ \textbf{PSNR (dB)}}  & \multicolumn{1}{c|}{ $\uparrow$ \textbf{SSIM (\%)}}   & $\downarrow$ \textbf{RMSE (\%)}   & \multicolumn{1}{c|}{ $\uparrow$ \textbf{PSNR (dB)}}  & \multicolumn{1}{c|}{ $\uparrow$ \textbf{SSIM (\%)}}   & $\downarrow$ \textbf{RMSE (\%)}   \\ \hline
\multicolumn{1}{|c|}{RED-CNN}    & \multicolumn{1}{c|}{37.09 ± 2.45} & \multicolumn{1}{c|}{93.01 ± 1.41} & 14.61 ± 0.46 & \multicolumn{1}{c|}{38.03 ± 2.41} & \multicolumn{1}{c|}{95.50 ± 1.81} & 13.05 ± 0.40 \\ \hline
\multicolumn{1}{|c|}{N2N}        & \multicolumn{1}{c|}{37.15 ± 2.52} & \multicolumn{1}{c|}{93.05 ± 1.72} & 14.50 ± 0.45 & \multicolumn{1}{c|}{39.47 ± 2.46} & \multicolumn{1}{c|}{96.26 ± 1.18} & 11.09 ± 0.32 \\ \hline
\multicolumn{1}{|c|}{EDCNN}      & \multicolumn{1}{c|}{36.50 ± 2.10} & \multicolumn{1}{c|}{92.56 ± 1.50} & 15.40 ± 0.41 & \multicolumn{1}{c|}{37.79 ± 2.42} & \multicolumn{1}{c|}{94.19 ± 2.66} & 13.65 ± 0.42 \\ \hline
\multicolumn{1}{|c|}{WGAN}       & \multicolumn{1}{c|}{35.91 ± 2.39} & \multicolumn{1}{c|}{91.84 ± 2.28} & 16.17 ± 0.47 & \multicolumn{1}{c|}{37.09 ± 1.86} & \multicolumn{1}{c|}{94.32 ± 2.18} & 13.92 ± 0.37 \\ \hline
\multicolumn{1}{|c|}{DU-GAN}     & \multicolumn{1}{c|}{35.35 ± 2.14} & \multicolumn{1}{c|}{90.65 ± 2.33} & 17.65 ± 0.47 & \multicolumn{1}{c|}{37.21 ± 2.33} & \multicolumn{1}{c|}{94.29 ± 3.68} & 14.49 ± 0.42 \\ \hline
\multicolumn{1}{|c|}{IDDPM}      & \multicolumn{1}{c|}{37.75 ± 2.60} & \multicolumn{1}{c|}{94.51 ± 1.73} & 13.23 ± 0.43 & \multicolumn{1}{c|}{39.34 ± 2.56} & \multicolumn{1}{c|}{96.12 ± 1.55} & 11.29 ± 0.36 \\ \hline
\multicolumn{1}{|c|}{UFormer}    & \multicolumn{1}{c|}{37.57 ± 2.62} & \multicolumn{1}{c|}{94.29 ± 1.50} & 13.88 ± 0.46 & \multicolumn{1}{c|}{39.38 ± 2.42} & \multicolumn{1}{c|}{96.18 ± 1.74} & 11.20 ± 0.33 \\ \hline
\multicolumn{1}{|c|}{LIT-Former} & \multicolumn{1}{c|}{36.50 ± 2.25} & \multicolumn{1}{c|}{92.71 ± 1.75} & 15.41 ± 0.44 & \multicolumn{1}{c|}{38.57 ± 2.36} & \multicolumn{1}{c|}{96.15 ± 1.57}  & 12.35 ± 0.37 \\ \hline
\multicolumn{1}{|c|}{ViMEDnet}   & \multicolumn{1}{c|}{37.50 ± 2.35} & \multicolumn{1}{c|}{93.52 ± 1.52} & 14.39 ± 0.40 & \multicolumn{1}{c|}{39.09 ± 2.43} & \multicolumn{1}{c|}{96.18 ± 1.70} & 11.62 ± 0.35 \\ \hline
\multicolumn{1}{|c|}{\textbf{DenoMamba}}  & \multicolumn{1}{c|}{\textbf{38.04 ± 2.20}} & \multicolumn{1}{c|}{\textbf{94.88 ± 1.57}} & \textbf{12.86 ± 0.38} & \multicolumn{1}{c|}{\textbf{39.72 ± 2.42}} & \multicolumn{1}{c|}{\textbf{96.33 ± 1.70}} & \textbf{10.76 ± 0.32} \\ \hline
\end{tabular}%
}
\label{tab:gen2}
\end{table*}

Representative denoised images recovered by competing methods are displayed in Fig. \ref{fig:ldct_1}. Among competing methods, convolutional baselines can alleviate local noise patterns in regions of homogeneous tissue signal, but they yield suboptimal depiction of detailed tissue structure that extend over longer distances, particularly near regions of heterogeneous tissue composition. Generative baselines typically yield a higher degree of visual sharpness in denoised images, albeit at the expense of elevated noise in recovered images that is particularly evident for adversarial models. Although contextually-sensitive baselines including ViMEDNet offer improved preservation of tissue structure across heterogeneous regions, they suffer from residual local noise patterns that can manifest as signal intensity fluctuations in homogeneous regions. In comparison, DenoMamba recovers high-quality CT images with more effective suppression of noise patterns, and accurate depiction of tissue structure and contrast. 

We also conducted experiments on the 10\%-dose dataset to assess competing methods in a relatively more challenging denoising task. Table \ref{tab:main2} lists performance metrics for competing methods on the test set. Corroborating the findings on the 25\%-dose dataset, we find that DenoMamba significantly outperforms all competing methods consistently across all examined settings (p$<$0.05). On average, DenoMamba achieves performance improvements of 1.9dB PSNR, 1.6\% SSIM, 2.5\% RMSE over convolutional baselines; 2.0dB PSNR, 1.8\% SSIM, 2.6\% RMSE over generative baselines, and 1.4dB PSNR, 1.5\% SSIM, 1.8\% RMSE over contextually-sensitive baselines. 

Representative denoised images recovered by competing methods are displayed in Fig. \ref{fig:ldct_2}. Note that prominent noise is apparent in LDCT images given the more aggressive dose reduction in 10\%-dose scans. Naturally, this elevates the difficulty of the LDCT denoising task as it becomes challenging to distinguish noise patterns from native variations in tissue signals. We observe that convolutional baselines can still offer reasonable suppression of local noise patterns in homogeneous regions, albeit this suppression comes at the expense of structural artifacts evident in regions of heterogeneous tissue composition. Meanwhile, generative baselines suffer from varying levels of noise amplification that can compromise structural accuracy particularly near tissue boundaries. Although contextually-sensitive baselines including ViMEDNet tend to improve depiction of tissue contrast over heterogeneous regions, they suffer from a degree of spatial blurring that can cause suboptimal depiction of fine tissue structures. Contrarily, DenoMamba offers high-fidelity depiction of detailed tissue structure in CT images and visibly improved suppression of noise. These results suggest that DenoMamba attains a more favorable balance between contextual sensitivity and local precision than competing methods, including ViMEDNet as a conventional SSM baseline. 

\begin{figure*}[t]
\centerline{\includegraphics[width=0.88\textwidth]{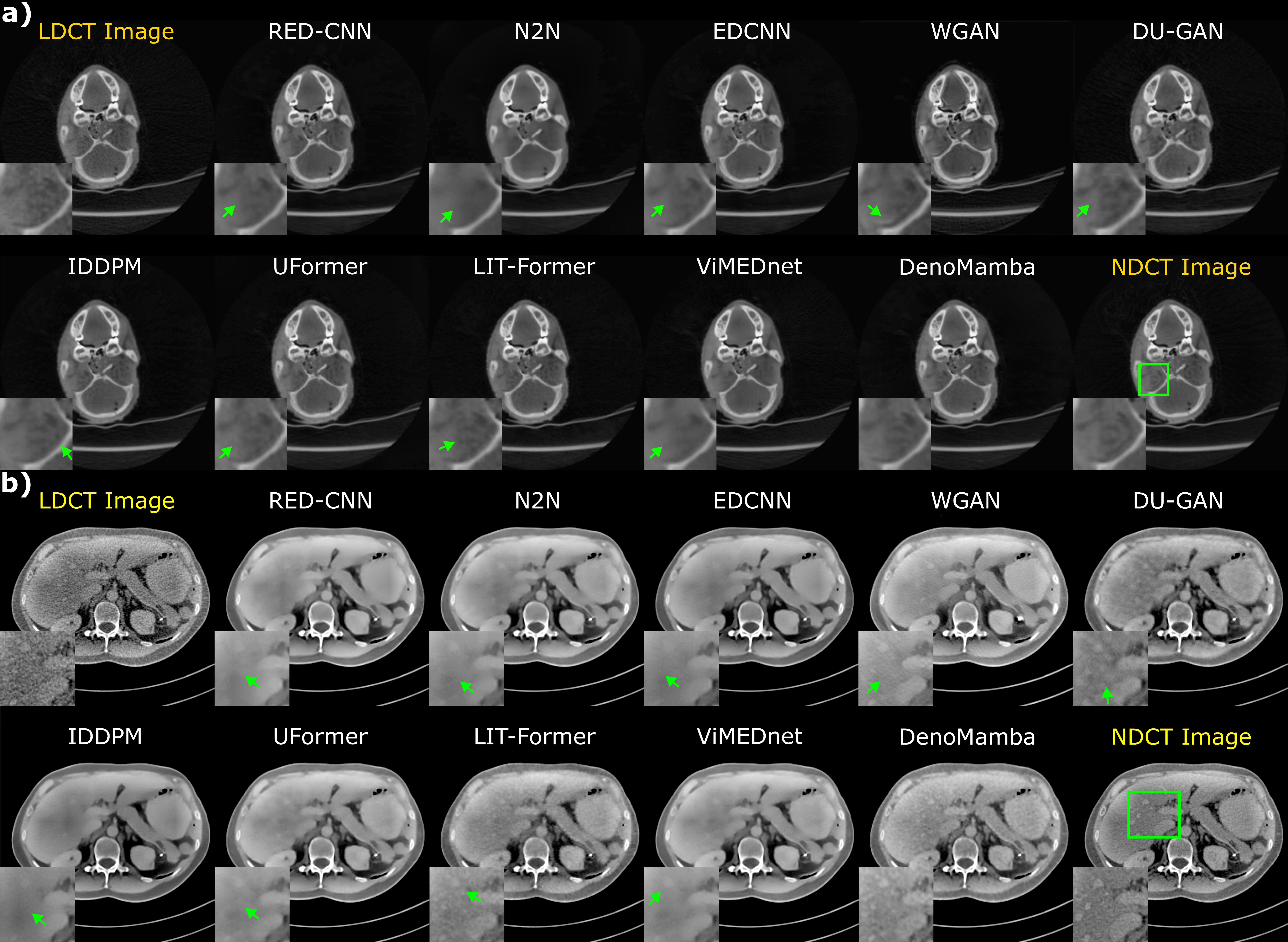}}
\caption{Denoising results for representative cross-sections from the experiments conducted to assess model generalization. \textbf{a)} Models trained on the 25\%-dose AAPM dataset were evaluated on the 10\%-dose Piglet CT dataset. \textbf{b)} For the AAPM dataset, models trained on 25\%-dose scans were evaluated on 10\%-dose scans. Display windows of \textbf{a)} [-400 1000] HU and \textbf{b)} [-250 450] HU are used.}
\label{fig:ldct_3}
\end{figure*}

\subsection{Generalization Performance}
Next, we conducted experiments to examine the generalizability of competing methods under domain shifts. First, we assessed denoising performance under shifts in the underlying data distribution for CT scans. For this purpose, models separately trained on the 25\%-dose and 10\%-dose AAPM datasets were independently tested on the 10\%-dose Piglet CT dataset. Table \ref{tab:gen1} lists performance metrics for competing methods. For both dose levels on which the models were trained, we find that DenoMamba significantly outperforms all competing methods in generalization across datasets (p$<$0.05). On average, DenoMamba achieves performance improvements of 1.3dB PSNR, 1.8\% SSIM, 3.5\% RMSE over convolutional baselines; 2.5dB PSNR, 4.5\% SSIM, 7.8\% RMSE over generative baselines, and 1.0dB PSNR, 1.5\%SSIM, 2.0\% RMSE over contextually-sensitive baselines. We also find that DenoMamba offers comparable levels of performance benefits over baselines in both examined settings, i.e., training on the 25\%-dose and training on the 10\%-dose AAPM scans. Yet, the absolute denoising performance of several competing methods including DenoMamba are moderately higher when trained on the 25\%-dose scans, even though the evaluations are conducted on the 10\%-dose Piglet CT scans. Through visual inspection, we confirmed that the 10\%-dose Piglet CT scans have more similar levels of noise perturbation to the 25\%-dose as opposed to 10\%-dose AAPM scans. Therefore, our findings are best attributed to the closer alignment of noise levels between training and test datasets, achieved when models are trained on the 25\%-dose AAPM scans. Representative images recovered by competing methods are depicted in Fig. \ref{fig:ldct_3}a. We observe that baseline models either suffer from over-smoothing manifested as spatial blurring (e.g., convolutional baselines, ViMEDNet) or from residual noise patterns manifested as structural artifacts (e.g., generative baselines, transformers) that can both compromise visibility of moderate variations in tissue contrast in denoised CT images. In comparison, DenoMamba recovers high-fidelity images with a closer appearance to reference NDCT images in terms of tissue structure and contrast. Collectively, these results indicate that DenoMamba shows a notable degree of robustness against shifts in the data distribution driven by native variations in anatomy and/or scanner hardware. 

We then assessed denoising performance under shifts in the level of dose reduction. To this end, models trained on 25\%-dose scans were tested on 10\%-dose scans, and models trained on 10\%-dose scans were tested on 25\%-dose scans in the AAPM dataset. Table \ref{tab:gen2} lists performance metrics for competing methods. For learning-based models, notable differences in image noise encountered between training and test sets can naturally induce performance losses. Yet, we find that DenoMamba significantly outperforms all competing methods in denosing performance (p$<$0.05), consistently in both shift directions (25\%$\rightarrow$10\%, 10\%$\rightarrow$25\%). On average across directions, DenoMamba achieves performance improvements of 1.2dB PSNR, 1.5\% SSIM, 1.9\% RMSE over convolutional baselines; 1.8dB PSNR, 2.0\% SSIM, 2.6\% RMSE over generative baselines, and 0.8dB PSNR, 0.8\% SSIM, 1.3\% RMSE over contextually-sensitive baselines. We also find that DenoMamba generally offers relatively higher levels of performance benefits over baselines in the shift direction of 25\%-dose$\rightarrow$10\%-dose versus 10\%-dose$\rightarrow$25\%-dose. This result implies that DenoMamba shows improved reliability against elevated task difficulty in the test set compared to baselines. Representative images recovered by competing methods are depicted in Fig. \ref{fig:ldct_3}b. High degrees of spatial blurring are apparent in convolutional baselines, DU-GAN, IDDPM, Uformer and ViMEDNet, which can be attributed to an overestimation of the noise level in LDCT images by the respective domain-transferred models. This spatial blurring yields suboptimal depiction of prominent vessel structures in abdominal images. Meanwhile, remaining methods including WGAN, LIT-Former and DenoMamba that are less amenable to spatial blurring show higher levels of residual noise. Among these methods, DenoMamba offers improved accuracy in depiction of important vascular structures evident in reference NDCT images, despite elevated levels of residual noise. Taken together, these results demonstrate that DenoMamba shows a degree of robustness against shifts in noise levels of CT scans to maintain its superior performance over baselines.

\begin{table}[t]
\centering
\caption{Performance of DenoMamba variants built by replacing SSM modules with vanilla transformers and image downsampling to 128$\times$128 (w ViT+down), with vanilla transformers and split processing of 128$\times$128 image patches (w ViT+patch), and with efficient transformers of linear complexity (w eff. ViT). Inference time and validation PSNR, SSIM, RMSE are listed for the 25\%-dose dataset.}
\setlength\tabcolsep{3pt}
\renewcommand{\arraystretch}{1.4}
\resizebox{\columnwidth}{!}{%
\begin{tabular}{c|c|ccc|}
\cline{2-5}
                                                                      & \multicolumn{1}{c|}{\textbf{Time (s)}} & \multicolumn{1}{c|}{ $\uparrow$ \textbf{PSNR (dB)}} & \multicolumn{1}{c|}{ $\uparrow$ \textbf{SSIM (\%)}} & $\downarrow$ \textbf{RMSE (\%)} \\ \hline
\multicolumn{1}{|c|}{w ViT+down}  & 0.18                                  &  \multicolumn{1}{c|}{38.79}     & \multicolumn{1}{c|}{92.26}    & 11.68    \\ \hline
\multicolumn{1}{|c|}{w ViT+patch} & 0.22                                  &  \multicolumn{1}{c|}{40.45}     & \multicolumn{1}{c|}{95.04}    & 9.93     \\ \hline
\multicolumn{1}{|c|}{w eff. ViT}                         & 0.10                                  & \multicolumn{1}{c|}{41.50}     & \multicolumn{1}{c|}{96.11}    & 8.97     \\ \hline
\multicolumn{1}{|c|}{\textbf{DenoMamba}}                                       & 0.15                                  & \multicolumn{1}{c|}{42.41}     & \multicolumn{1}{c|}{96.75}    & 8.31     \\ \hline
\end{tabular}%
}
\label{tab:ab1}
\end{table}

\subsection{Ablation Studies}
We conducted a systematic set of ablation studies to examine the importance of key building elements and design parameters in DenoMamba. First, we assessed the efficacy of SSM modules in DenoMamba for capturing contextual representations in comparison to transformer modules. Note that vanilla transformers (ViT) induce quadratic complexity with respect to sequence length \cite{vaswani2017}, which prohibited the use of ViT modules at the original image resolution given memory limitations on GPUs employed in the current study. Thus, transformer-based variants were formed by adopting several different strategies to mitigate complexity. A `w ViT+down' variant was formed by replacing the SSM modules with ViT modules, and spatially downsampling images to a 128$\times$128 size \cite{trans_unet}. A `w ViT+patch' variant was formed by replacing the SSM modules with ViT modules, splitting each image into a set of four 128$\times$128 patches, and processing separate patches individually \cite{trans_review1}. A `w eff. ViT' variant was formed by adopting an efficient transformer module of linear complexity based on transposed attention \cite{restormer}. Table \ref{tab:ab1} lists performance metrics for DenoMamba and transformer-based variants on the 25\%-dose dataset, along with inference times per slice. DenoMamba outperforms all variant models in performance metrics (p$<$0.05). We find that DenoMamba achieves relatively stronger performance benefits over  `w ViT+down' and `w ViT+patch', along with shorter inference times. These results suggest that compromising image resolution or field-of-view in transformer modules that inherently restricts spatial precision causes notable losses in image quality. While `w eff. ViT' offers the shortest inference time among all models, DenoMamba still attains significant improvements in image quality over this efficient transformer variant, suggesting that SSM modules have higher efficacy in learning contextual representations.

\begin{table}[t]
\centering
\caption{Performance of DenoMamba variants built by ablating the channel SSM module (w/o cha. SSM), the spatial SSM module (w/o spa. SSM), the CFM module (w/o CFM), the gated convolution network to extract latent features in the channel SSM module (w/o GCN), and the identity path that relays input features to the CFM module (w/o Iden.).}
\resizebox{0.9\columnwidth}{!}{%
\begin{tabular}{|c|c|c|c|}
\cline{2-4}
\multicolumn{1}{c|}{ }   & $\uparrow$ \textbf{PSNR (dB)}  & $\uparrow$ \textbf{SSIM (\%)} & $\downarrow$ \textbf{RMSE (\%)}  \\ \hline
w/o spa. SSM                                                & 42.10          & 96.71          & 8.62           \\ \hline
w/o cha. SSM                                                & 41.93          & 96.69          & 8.76           \\ \hline
w/o CFM                                                     & 42.27          & 96.73          & 8.41           \\ \hline
w/o GCN                                                  & 42.31          & 96.73          & 8.39           \\ \hline
w/o Iden.                                                     & 42.05          & 96.46          & 8.70           \\ \hline
\textbf{DenoMamba}                                                    & 42.41          & 96.75          & 8.31           \\ \hline
\end{tabular}%
}
\label{tab:ab2}
\end{table}

\begin{table}[t]
\centering
\caption{Performance of DenoMamba variants built by varying the number of encoder-decoder stages $K$, the number of feature channels $C$, and the configuration for the number of FuseSSM blocks across stages $E-D$.}
\resizebox{0.9\columnwidth}{!}{%
\begin{tabular}{cc|c|c|c|}
\cline{3-5}
\multirow{1}{*}{} & \multirow{1}{*}{} &                 $\uparrow$ \textbf{PSNR (dB)}  & $\uparrow$ \textbf{SSIM (\%)} & $\downarrow$ \textbf{RMSE (\%)} \\ \hline
\multicolumn{1}{|c|}{\multirow{3}{*}{$K$}} & 3                                                                                     & \multicolumn{1}{c|}{42.30} & \multicolumn{1}{c|}{96.73} & 8.41 \\ \cline{2-5} 
\multicolumn{1}{|c|}{}                      & 4                                                                                     & \multicolumn{1}{c|}{42.41} & \multicolumn{1}{c|}{96.75} & 8.31 \\ \cline{2-5} 
\multicolumn{1}{|c|}{}                      & 5                                                                                      & \multicolumn{1}{c|}{42.28} & \multicolumn{1}{c|}{96.73} & 8.47 \\ \hline
\multicolumn{1}{|c|}{\multirow{3}{*}{$C$}}  & 32                                                                                      & \multicolumn{1}{c|}{42.25} & \multicolumn{1}{c|}{96.71} & 8.44 \\ \cline{2-5} 
\multicolumn{1}{|c|}{}                      & 48                                                                                    & \multicolumn{1}{c|}{42.41} & \multicolumn{1}{c|}{96.75} & 8.31 \\ \cline{2-5} 
\multicolumn{1}{|c|}{}                      & 56                                                                                   & \multicolumn{1}{c|}{42.23} & \multicolumn{1}{c|}{96.72} & 8.45 \\ \hline
\multicolumn{1}{|c|}{\multirow{3}{*}{$E-D$}}  & \underline{1}                                                                                 & \multicolumn{1}{c|}{42.26} & \multicolumn{1}{c|}{96.72} & 8.39 \\ \cline{2-5} 
\multicolumn{1}{|c|}{}                      & \underline{2}                                                                               & \multicolumn{1}{c|}{42.41} & \multicolumn{1}{c|}{96.75} & 8.31 \\ \cline{2-5} 
\multicolumn{1}{|c|}{}                      & \underline{3}                                                          & \multicolumn{1}{c|}{42.39} & \multicolumn{1}{c|}{96.75} & 8.33 \\ \hline
\end{tabular}%
}
\label{tab:ab3}
\end{table}

We then assessed the influence of individual modules in DenoMamba on denoising performance. Several ablated variants were formed for this purpose. A `w/o spa. SSM' variant was formed by ablating the spatial SSM module in FuseSSM blocks. A `w/o cha. SSM' variant was formed by ablating the channel SSM module in FuseSSM blocks. A `w/o CFM' variant was formed by replacing the channel fusion module in FuseSSM blocks with a simple element-wise addition operator to combine contextual features from spatial/channel SSM modules with input features. A `w/o GCN' variant was formed by ablating the gated convolutional network in channel SSM modules that extracts latent contextual features across the channel dimension. A `w/o Iden.' variant was formed by ablating the identity propagation path in FuseSSM blocks that relays input features to the CFM module. Table \ref{tab:ab2} lists performance metrics for DenoMamba and ablated variants on the 25\%-dose dataset, along with the number of model parameters. We find that DenoMamba outperforms all ablated variants (p$<$0.05). Higher performance of DenoMamba over the `w/o spa. SSM', `w/o cha. SSM', and `w/o Iden.' variants indicate that contextual features in spatial and channel dimensions along with lower-level spatial features effectively contribute to LDCT denoising performance. Note that low-level input features can be propagated across FuseSSM blocks in multiple ways, including the identity propagation path feeding into the CFM module where input and contextual features are subjected to nonlinear convolutional fusion, as well as the residual connections in channel and spatial SSM modules that additively fuse the input and contextual features. Taken together, higher performance of DenoMamba against the `w/o Iden.' variant that removes the identity path, and against the `w/o CFM' variant that additively combines feature sets indicate that nonlinear convolutional fusion better preserves low-level representations of CT images than additive fusion via residual connections. 

Lastly, we assessed the influence of the number of encoder-decoder stages $K$, the number of initial feature channels at the first encoder stage $C$ (note that the number of feature channels in remaining stages scale proportionately with $C$), and the numbers of FuseSSM blocks cascaded across individual encoder-decoder stages $E-D$ (i.e., the number of FuseSSM blocks across $K$ encoder and $K$ decoder stages). In general, prescribing higher values for these design parameters increases model complexity. As learning-based models are subject to an intrinsic trade-off between allowed degrees of freedom versus learning efficacy, we wanted to examine whether the selected design parameters for DenoMamba offer a favorable compromise. For this purpose, variant models were built by separately varying the values of $K$, $C$, and $R$ while remaining parameters were kept fixed. Specifically, we varied $K$ in \{3, 4, 5\}; $C$ in \{32, 48, 56\}; and $E-D$ in \{\underline{1}: [2, 3, 3, 4] - [3, 3, 2, 1], \underline{2}: [4, 6, 6, 8] - [6, 6, 4, 2], \underline{3}: [6, 9, 9, 12] - [9, 9, 6, 3]\}. Table \ref{tab:ab3} lists performance metrics of DenoMamba variants on the 25\%-dose dataset. We find that variants for $K=4$, $C=48$, and $E-D=\underline{2}$ yield near-optimal performance, validating the proposed selection of design parameters.

\section{Discussion}
In this study, we introduced a novel denoising method to recover high-quality NDCT images from noisy LDCT images. Previous CNN models offer a high degree of local precision, albeit they are relatively insensitive to long-range relationships between distant anatomical regions in medical images \cite{slater}. While transformer models can address this limitation by leveraging the long-range contextual sensitivity of self-attention operators, they inherently suffer from quadratic model complexity with respect to sequence length \cite{kodali2018}. Meanwhile, common approaches to mitigate this complexity result in inevitable losses in spatial precision \cite{trans_review2}. Differently from these previous models, DenoMamba employs novel FuseSSM blocks to capture contextual features via state-space modeling across spatial and channel dimensions, without compromising local precision. Our demonstrations indicate that DenoMamba achieves superior performance in LDCT denoising against state-of-the-art CNN, transformer and SSM methods, with apparent quantitative and qualitative benefits in recovered CT images.

Several technical limitations can be addressed in order to further boost the performance and practicality of DenoMamba. A first line of improvements concerns the nature of denoising tasks targeted during model training. Here, a separate model was trained for LDCT denoising at each reduction level for radiation dose to maintain high performance. Note that this may lower practicality if highly variable reduction levels are expected to be administered in practice. In those cases, DenoMamba can be trained on LDCT images at varying reduction levels, and model specialization to specific radiation doses could be enhanced by adaptive normalization approaches on feature maps \cite{pFLSynth,cbm_2}. This could improve practicality by building a unified model that can be deployed at various dose reduction levels. 

A second line of improvements concerns the datasets on which DenoMamba is trained to perform LDCT denoising. Here, we performed supervised learning relying on the availability of paired LDCT-NDCT images from the same set of subjects \cite{pgan}. Note that, in practice, the curation of such paired datasets can be challenging as it would require repeated CT scans on a given subject at separate radiation doses. In cases where the amount of paired training data that can be collected is limited, a large training set can be curated by instead adopting cycle-consistent learning procedures on unpaired sets of LDCT and NDCT images \cite{syndiff}, or self-supervised learning procedures to train models directly on LDCT measurements \cite{ssdiffrecon,9577596}.

A third line of improvements concerns the loss terms employed to train DenoMamba. Here, we utilized a simple pixel-wise loss term based on mean absolute error, since we observed that this pixel-wise loss offered effective learning of LDCT denoising models on the examined datasets. That said, it might be possible to attain further improvements in recovered image quality by using more sophisticated loss terms including adversarial, score-based or cross-entropy losses \cite{IDDPM,FDB}. Particularly within the context of score-based methods that involve iterative sampling procedures, the long-range contextual sensitivity of DenoMamba combined with task-driven bridge formulations might offer benefits over conventional denoising diffusion models based on CNN backbones \cite{gao2023corediff,Du2024PMB,dar2022adaptive}. Further work is warranted for a systematic evaluation of the utility of various loss functions on the performance and reliability of DenoMamba.

\section{Conclusion}
Here we introduced a novel fused state-space model (SSM) for recovery of high-quality images from noisy LDCT scans. The proposed DenoMamba model leverages an hourglass architecture implemented with novel FuseSSM blocks. Each FuseSSM block extracts contextual features across spatial and channel dimensions via spatial and channel SSM modules, respectively, and performs fusion of contextual and low-level input features via a CFM module. This design enables DenoMamba to leverage contextual relationships in LDCT images without compromising local precision, and thereby to offer superior performance against state-of-the-art LDCT denoising methods. Therefore, DenoMamba holds great promise for performant LDCT image denoising.

\bibliographystyle{IEEEtran.bst}
\bibliography{refs}

\end{document}